\documentclass[conference]{IEEEtran}
\ifCLASSINFOpdf
\else
\fi
\usepackage{subfigure}
\usepackage{fancybox}
{\begin{center}\vspace{0mm}\noindent\begin{Sbox}\begin{minipage}{0.97\columnwidth}}%
{\end{minipage}\end{Sbox}\fbox{\TheSbox}\end{center}\vspace{0mm}}

\usepackage{caption}
\usepackage{hyperref}
\usepackage{booktabs}
\usepackage{graphicx} 
\usepackage{tabularx}

\usepackage{tcolorbox}

\def\BibTeX{{\rm B\kern-.05em{\sc i\kern-.025em b}\kern-.08em
T\kern-.1667em\lower.7ex\hbox{E}\kern-.125emX}}

\tcbuselibrary{skins}
\newtcolorbox{rqbox}[1]
{
  before skip=1em, 
  after skip=1em, 
  colframe = black!60,
  colback  = black!10,
  coltitle = black!90,  
  title    = \textbf{#1},
  hbox boxed title,
  enhanced,
  attach boxed title to top center={yshift=-3mm,yshifttext=-1mm},
  boxed title style={size=small, colback=black!30}
} 

\begin{document}
%

\title{Evidence is All We Need: Do Self-Admitted Technical Debts Impact Method-Level Maintenance?}

\author{
\IEEEauthorblockN{1\textsuperscript{st} Shaiful Chowdhury}
\IEEEauthorblockA{\textit{SQM Research Lab, Computer Science}\\
\textit{University of Manitoba}\\
Winnipeg, Canada\\
Email: shaiful.chowdhury@umanitoba.ca}
\and 
\IEEEauthorblockN{2\textsuperscript{nd} Hisham Kidwai}
\IEEEauthorblockA{\textit{SQM Research Lab, Computer Science}\\
\textit{University of Manitoba}\\
Winnipeg, Canada\\
Email: kidwaih1@myumanitoba.ca}
\and
\IEEEauthorblockN{3\textsuperscript{rd} Muhammad Asaduzzaman}
\IEEEauthorblockA{\textit{School of Computer Science}\\
\textit{University of Windsor}\\
Windsor, Canada\\
Email: masaduzz@uwindsor.ca
}
}


%


\maketitle

\begin{abstract}

Self-Admitted Technical Debt (SATD) refers to the phenomenon where developers explicitly acknowledge technical debt through comments in the source code. While considerable research has focused on detecting and addressing SATD, its true impact on software maintenance remains underexplored. The few studies that have examined this critical aspect have not provided concrete evidence linking SATD to negative effects on software maintenance. These studies, however, focused only on file- or class-level code granularity. This paper aims to empirically investigate the influence of SATD on various facets of software maintenance at the method level. We assess SATD's effects on code quality, bug susceptibility, change frequency, and the time practitioners typically take to resolve SATD.

By analyzing a dataset of 774,051 methods from 49 open-source projects, we discovered that methods containing SATD are not only larger and more complex but also exhibit lower readability and a higher tendency for bugs and changes. We also found that SATD often remains unresolved for extended periods, adversely affecting code quality and maintainability.
Our results provide empirical evidence highlighting the necessity of early identification, resource allocation, and proactive management of SATD to mitigate its long-term impacts on software quality and maintenance costs.

\end{abstract}



\section{Introduction}
\label{intro}

Software maintenance is a crucial component of the software development lifecycle, involving activities such as bug fixing, feature enhancement, performance improvement, and perfective changes. The costs associated with software maintenance have long been a concern~\cite{Kafura:1987}, often surpassing the original development costs~\cite{Borstler:2016} and accounting for approximately 70\% of the total software expenditure~\cite{Carr:2002}. As a result, numerous studies have focused on identifying and predicting components of code that are prone to maintenance issues, enabling proactive measures to alleviate future maintenance challenges~\cite{Chowdhury:2024:promise, Pascarella:2020, Romano:2011, Spadini:2018, Zhou:2010, Chowdhury:2022-deb}.

Technical debt~\cite{de2019tracy,Potdar:2014,esfandiari2023exploratory} is a metaphor for compromises made in technical quality that yield short-term benefits but may adversely affect software quality in the long term. A common form of technical debt is self-admitted technical debt (SATD), where developers leave comments in the source code to acknowledge suboptimal solutions, such as temporary workarounds, while signaling the need for future improvements~\cite{Potdar:2014,Li:2015}. SATD is considered detrimental to future maintenance, prompting extensive research on both its detection~\cite{Liu:2018, ren2019neural, maldonado2015detecting, li2023automatic, guo2021far, sharma2022self} and removal~\cite{zampetti2020automatically, zampetti2018self, maldonado2017empirical}. SATD is deemed so detrimental that during code review processes, code containing SATD is accepted significantly less frequently than code that is free of such debt~\cite{kashiwa2022empirical}. Unfortunately, making optimizations to remove SATD often requires significant effort~\cite{li4724886presti}.

Despite the substantial research and effort dedicated to SATD detection and removal based on the assumption that \emph{SATD is detrimental to maintenance}, it is surprising that these claims lack substantial validation through empirical evidence. The few studies that have investigated the impact of SATD on software quality or maintenance found no concrete evidence to support its adverse effects.  For instance, a study by Wehaibi \textit{et al.}~\cite{wehaibi2016examining} identified no clear correlation between defects and SATD. According to the authors, \emph{``In some of the studied projects, self-admitted technical debt files have
more bug-fixing changes, while in other projects, files without it had more defects.''} Similarly, Potdar and Shihab~\cite{Potdar:2014} evaluated the relationship between code quality and SATD and found that SATD did not affect any of the studied complexity metrics related to quality. This unexpected observation was also noted by Bavota and Russo~\cite{bavota2016large}. Interestingly, although it is hypothesized that practitioners incur SATD due to time pressure, no empirical evidence was found to support this hypothesis when tested~\cite{Potdar:2014, sierra2019survey}.

This raises a crucial question: \emph{are we investing time in developing methods for detecting and resolving SATD without any concrete evidence supporting their negative impacts?} As a scientific community, we recognize the importance of intuition in generating hypotheses and guiding exploration. However, our decision-making processes must be grounded in empirical data and rigorous evidence, rather than relying solely on intuition or anecdotal observations~\cite{kitchenham2004evidence}.  A compelling example is the work of Inozemtseva and Holmes~\cite{Inozemtseva:2014}, who critically examined the long-held belief that code coverage is a reliable indicator of test suite effectiveness. After a thorough evaluation, the authors found that the correlation between coverage and effectiveness is not strong. This landmark paper, which won the 2024 ICSE Most Influential Paper Award, shifted the paradigm by promoting more nuanced testing strategies by balancing coverage with other relevant metrics in assessing test suite quality. 

In this paper, we critically evaluate the prevailing belief regarding the negative impact of SATD, and pose the question, \emph{is SATD truly harmful to software quality and maintenance?} Our analysis focuses on method-level granularity, which is crucial because previous studies~\cite{Potdar:2014, bavota2016large, wehaibi2016examining}---which found no clear evidence of SATD's negative impact---were conducted at the class or file level. Our motivation to study SATD at the method level is driven by three key observations:

\begin{itemize} 
\item Practitioners find method-level maintenance tasks significantly easier than those at the file or class level~\cite{Grund:2021, Pascarella:2020}, as identifying and resolving issues at the method level typically requires much less time—methods are generally smaller than files or classes. 
\item Numerous studies have demonstrated no correlation between file or class-level internal software metrics and maintenance burdens~\cite{Gil:2017, Emam:2001, TSE:2013}, probably due to the lack of variability in measurements, as classes and files are typically large. However, these metrics have proven to be valuable indicators of maintenance when evaluated at the method level~\cite{Landman:2014, Chowdhury:2022-deb, Chowdhury:2024:promise}---methods are generally much smaller than classes and files.
\item Through various transformations, a method containing SATD can be relocated to a different file~\cite{Grund:2021, jodavi2022accurate}. As a result, tracking SATD at the class or file level, as done in previous studies~\cite{Potdar:2014, bavota2016large, wehaibi2016examining}, may lead to inaccuracies. 
\end{itemize}




To gain an accurate understanding of the method-level maintenance impact of SATD, we analyze 774,051 Java methods from 49 popular open-source projects. Our investigation focuses on how SATD affects code quality, bug-proneness, and change-proneness of source code methods. Our contributions are based on the following four research questions: 

\textbf{RQ1: \textit{Do methods with SATD impact code quality?}}
We evaluate 14 popular method-level code metrics, and the results are consistent across all metrics. 
In general, our analysis indicates that methods containing SATD exhibit significantly lower code quality, characterized by reduced readability and increased size and complexity. 

\textbf{RQ2: \textit{Do methods with SATD impact code change-proneness?}}
We demonstrate that methods with SATD are more change-prone, undergoing more frequent revisions than those without SATD. This increased change-proneness indicates that SATD is a strong predictor of future maintenance challenges, leading to higher maintenance costs.

\textbf{RQ3: \textit{Do methods with SATD impact code bug-proneness?}}
Our study reveals that methods with SATD are more likely to be associated with bugs, as these methods are associated more strongly with bug-fix commits than their non-SATD counterparts. This highlights the critical importance of addressing SATD to enhance both maintainability and reliability.

\textbf{RQ4: \textit{How long does it take to resolve SATD?}} Given the empirical evidence of the negative impact of SATD on maintenance, we aim to understand the duration required to eliminate SATD from software systems. Unlike previous studies, we approach this question with a more precise change history of SATD, facilitated by recently developed method history tracking tools, such as CodeShovel~\cite{Grund:2021}. Our findings reveal that a major portion of SATDs ($>60\%$) is never removed from the software systems. Even when SATDs are resolved, they often persist for long periods, with many remaining unresolved for over a year. Notably, we show that 20\% of SATDs are resolved after 1,000 days, a very concerning observation not reported in earlier research.


In summary, with empirical evidence, our findings underscore the detrimental effect of SATD on software maintainability, highlighting the need for early detection and remediation. To enable replication, we share our dataset publicly\footnote{https://zenodo.org/records/14058483}.

\section{Related Work and Motivation}
\label{rlwork}

In this section, we review previous studies on SATD detection and removal, as well as maintenance indicators, to illustrate how these studies have inspired this paper.
\subsection{SATD Detection and Removal}
Potdar and Shihab~\cite{Potdar:2014} created a dataset of SATD and non-SATD by manually analyzing 101,762 comments from four large open-source projects. This substantial labeled dataset is crucial for developing models to automatically detect SATD. The authors also identified 62 comment patterns that can be leveraged for automatic SATD detection. Their study has inspired numerous investigations focused on SATD detection using various machine learning models (e.g.,~\cite{Maldonado:2017, Liu:2018, sharma2022self, ren2019neural, maldonado2015detecting}). In contrast, another stream of research has concentrated on the removal of SATD~\cite{maldonado2017empirical, zampetti2018self, zampetti2020automatically, iammarino2021empirical, liu2021exploratory, mastropaolo2023towards}. However, these studies have largely operated under the assumption that SATD is detrimental to software maintenance. 

\subsection{Maintenace Impact of SATD}
Surprisingly, only a limited number of studies have investigated the true maintenance impact of SATD. 
Zampetti \textit{et al.}~\cite{zampetti2017recommending} utilized different code metrics and warnings from different static analysis-based tools to identify when a design-specific debt should become SATD. While they identified a relationship between code metrics and design debt, they revealed that even in the best-performing scenario, the recall for predicting design debt was only 55\%. 
Potdar and Shihab~\cite{Potdar:2014} explored the effects of SATD on code quality by analyzing three key metrics: fan-in, fan-out, and cyclomatic complexity. Their findings revealed weak correlations between these metrics and SATD, with some relationships lacking statistical significance. Bavota and Russo~\cite{bavota2016large} conducted a replication study involving 159 software projects, building on Potdar and Shihab’s work. They considered factors such as coupling, complexity, and code readability, yet found no correlation between code quality and SATD. 

Similarly, Wehaibi \textit{et al.}~\cite{wehaibi2016examining} concluded that “\emph{there is no clear trend when it comes to defects and self-admitted technical debt}.” Notably, they observed that changes associated with SATD tend to be more complex and are likely to introduce more future bugs compared to non-SATD changes. This raises an intriguing question: \emph{If SATDs themselves are not inherently harmful, but addressing them may lead to greater issues, should our focus be on detecting and eliminating SATDs?}

\subsection{Code Metrics}
Internal software metrics, often referred to as code metrics, are widely used to estimate and predict external software metrics, such as bug-proneness and change-proneness~\cite{Pascarella:2020, Chowdhury:2024:promise, Gil:2017, Landman:2014, Shin:2011, Spadini:2018}. Unlike external metrics, internal metrics are easy to measure and can serve as reflective indicators of software quality and maintenance throughout the software development lifecycle. However, the effectiveness of code metrics in understanding maintenance has been a subject of extensive debate~\cite{Chowdhury:2022-deb}. For instance, Gil \textit{et al.}~\cite{Gil:2017} found that, with the exception of \emph{size}, code metrics are not particularly useful for building models that predict future maintenance burdens. This finding aligns with other similar studies~\cite{Emam:2001, TSE:2013}. Conversely, Landman \textit{et al.}~\cite{Landman:2014} noted that the limited usefulness of code metrics primarily applies at the file or class level. When analyzing metrics at the method level, however, metrics like McCabe's cyclomatic complexity~\cite{McCabe:1976} can serve as effective predictors of maintenance challenges. This has been further validated by multiple studies~\cite{Mo:2022, Chowdhury:2024:promise, Chowdhury:2022-deb}. \emph{These findings motivated our investigation into the code quality impact of SATD at the method-level granularity (RQ1).}
\subsection{Bug- and Change-proneness}
While the relationship between code metrics and maintenance has been debated, there is a consensus that change-proneness and bug-proneness of code components are among the most reliable indicators of software maintenance~\cite{Pascarella:2020, Chowdhury:2024:promise, Shippey:2016, Romano:2011, Moser:2008, Gil:2017, Khomh:2012, Palomba:2017}. Code components that exhibit higher bug-proneness can lead to significant future burdens, affecting both revenue and reputation. Similarly, well-designed code should require minimal revisions, in line with the \emph{open-closed principle}—software should be open for extension but closed for modification~\cite{ng2006toward}. \emph{Consequently, we compare the change- and bug-proneness of SATD methods with that of NOT-SATD methods (RQ2 and RQ3).} 

\subsection{SATD Removal Time}

Multiple studies have examined the duration it takes to remove SATDs from software projects. Potdar and Shihab~\cite{Potdar:2014} noted that up to 75\% of SATDs were never removed within the projects they studied. In some cases, only about 37\% of SATDs were addressed by the next release. This indicates that SATDs can persist in software projects for extended periods. This finding was further supported by Maldonado \textit{et al.}~\cite{maldonado2017empirical}, who reported that it may take 18 to 172 days (on median) to remove SATDs. Conversely, Liu \textit{et al.}~\cite{liu2021exploratory} found that SATDs are generally removed more quickly than Maldonado \textit{et al.} suggested. However, these studies did not effectively track SATD, particularly when methods containing SATD were moved or renamed. Generally, previous approaches treated SATD as removed when the containing method or file could no longer be found. With the advent of advanced method tracing tools like CodeShovel~\cite{Grund:2021} and CodeTracker~\cite{jodavi2022accurate}, we can now accurately trace methods even after they have been renamed or relocated.  
\emph{Therefore, with a new more accurate method-level dataset, we aim to re-investigate this issue: what percentage of SATDs are typically removed, and how long do the SATDs generally last in software projects (RQ4)?}

\section{Methodology}
\label{Methodology}

In this section, we outline our project selection criteria, methods for collecting change history, and SATD detection. We also cover the statistical tests utilized in our study. To enhance readability, methods unique to specific questions are addressed in their respective sections in Section~\ref{results}.

\subsection{Project Selection}
\label{subsec:project-selection-methodology}

To explore the influence of SATD on method-level maintenance, we selected a range of open-source software projects from GitHub\footnote{\href{https://github.com}{https://github.com}}, a leading platform for mining software repositories. Previous MSR-based research has typically taken one of two approaches: aggregated analysis~\cite{Gil:2017, Pascarella:2020, Spadini:2018} or individual project analysis~\cite{Kafura:1987, Romano:2011, Shin:2011, Zhou:2010}. In aggregated analysis, the results are presented as a summary of all projects combined, rather than individually for each project. While this approach can be useful, it is vulnerable to the influence of outliers, potentially skewing findings. On the other hand, individual project analysis provides a more granular view, and we can observe how SATD affects different projects. This approach, however, is prone to issues such as selection bias or publication bias~\cite{Ralph:2018, Gil:2017}.

To address these limitations, we conducted our study using both approaches on a selected set of 49 open-source Java projects---the tools we used, e.g., CodeShovel, are compatible only with Java, which limited our ability to analyze projects based on other programming languages. We present results aggregated across these projects, in addition to providing more detailed insights from individual project analyses. Also, rather than selecting projects at random, we selected all the projects used in five MSR-based studies~\cite{Gil:2017, Grund:2021, Palomba:2017, Ray:2016, Spadini:2018}, ensuring a broad and representative dataset. This approach allows us to draw more robust and widely applicable conclusions about the role of SATD in open-source software maintenance. Our analysis begins with a total of 774,051 open-source methods, as detailed in Table \ref{tab:49-source-methods}.

\begin{table*}[htbp]
\centering
\caption{We selected 49 projects for our study. These projects are popular within the open-source community, as indicated by their large number of contributors and stars. The percentage of SATD methods is also presented for each project.}
\label{tab:49-source-methods}
\begin{tabular}{lrrrrr}
\toprule
\textbf{Repository} & \textbf{\# Methods} & \textbf{\# SATD Methods (\%)} & \textbf{\# Contributors} & \textbf{\# Stars} & \textbf{Snapshot} \\
\midrule
hadoop&70081&1224 (1.75)&592&14500&\texttt{4c5cd7}\\
elasticsearch&62190&1372 (2.21)&1932&68400&\texttt{92be38}\\
flink&38081&5111 (13.42)&908&16764&\texttt{261e72}\\
lucene-solr&37133&1681 (4.53)&234&4400&\texttt{b457c2}\\
presto&36715&816 (2.22)&734&15700&\texttt{bb20eb}\\
docx4j&36514&2452 (6.72)&37&1620&\texttt{36c378}\\
hbase&36274&5853 (16.14)&463&5200&\texttt{3bd542}\\
intellij-community&35950&767 (2.13)&1035&16800&\texttt{cdf2ef}\\
weka&35639&513 (1.44)&2&323&\texttt{a22631}\\
hazelcast&35265&426 (1.21)&340&6000&\texttt{a59ad4}\\
spring-framework&26634&339 (1.27)&544&43641&\texttt{1984cf}\\
hibernate-orm&24800&930 (3.75)&420&4678&\texttt{2c12ca}\\
eclipseJdt&22124&1061 (4.80)&133&146&\texttt{475591}\\
guava&20757&491 (2.37)&264&41775&\texttt{e35207}\\
sonarqube&20627&249 (1.21)&137&5937&\texttt{6b806e}\\
jclouds&20358&182 (0.89)&231&384&\texttt{7af4d8}\\
wildfly&19665&645 (3.28)&343&2593&\texttt{f21f5d}\\
netty&16908&316 (1.87)&526&27185&\texttt{662e0b}\\
cassandra&15953&2217 (13.90)&447&8600&\texttt{7cdad3}\\
argouml&12755&3414 (26.77)&4&238&\texttt{fcbe6c}\\
jetty&10645&422 (3.96)&159&3160&\texttt{fc5dd8}\\
voldemort&10601&189 (1.78)&65&2475&\texttt{a7dbde}\\
spring-boot&10374&29 (0.28)&828&56378&\texttt{199cea}\\
wicket&10058&243 (2.42)&83&570&\texttt{e3f370}\\
ant&9781&187 (1.91)&58&303&\texttt{1ce1cc}\\
jgit&9548&208 (2.18)&140&983&\texttt{855842}\\
mongo-java-driver&9467&109 (1.15)&151&2422&\texttt{8ab109}\\
pmd&8992&282 (3.14)&229&3482&\texttt{d115ca}\\
xerces2-j&8153&289 (3.54)&3&38&\texttt{cf0c51}\\
RxJava&8145&145 (1.78)&278&44922&\texttt{880eed}\\
openmrs-core&6066&184 (3.03)&371&1017&\texttt{c5928a}\\
javaparser&5862&110 (1.88)&157&3771&\texttt{8f25c4}\\
hibernate-search&5345&212 (3.97)&58&390&\texttt{5b7780}\\
titan&4590&103 (2.24)&34&5135&\texttt{ee226e}\\
facebook-android-sdk&3759&69 (1.84)&124&6100&\texttt{fb1b91}\\
checkstyle&3340&89 (2.66)&287&6115&\texttt{164a75}\\
commons-lang&2948&58 (1.97)&160&2138&\texttt{f69235}\\
lombok&2684&78 (2.91)&114&10391&\texttt{4fdcdd}\\
atmosphere&2659&77 (2.90)&111&3524&\texttt{fadfb0}\\
jna&2636&54 (2.05)&143&6655&\texttt{b8443b}\\
Essentials&2390&52 (2.18)&215&1086&\texttt{d36d80}\\
junit5&2085&34 (1.63)&167&4691&\texttt{be2aa2}\\
hector&1958&53 (2.71)&71&648&\texttt{a302e6}\\
okhttp&1953&144 (7.37)&236&40428&\texttt{5224f3}\\
mockito&1498&107 (7.14)&21&12043&\texttt{77562}\\
cucumber-jvm&1146&26 (2.27)&226&2291&\texttt{b57b92}\\
commons-io&1145&23 (2.01)&76&787&\texttt{11f0ab}\\
vraptor4&926&9 (0.97)&48&343&\texttt{593ce9}\\
junit4&874&67 (7.67)&151&8160&\texttt{50a285}\\
\midrule
\textbf{Total} & \textbf{774051} & \textbf{33711 (4.36)}& & \\
\bottomrule
\end{tabular}
\end{table*}

\subsection{Tracking Change History}
\label{subsec:change-history-collection}
Tracking changes at the method level is inherently difficult due to factors such as renaming, reparameterizing, or relocating methods across different files~\cite{Grund:2021}. To overcome this challenge, we utilized CodeShovel~\cite{Grund:2021}, a tool designed to trace method histories even in the face of these transformations. In both precision and recall, CodeShovel outperformed other tools including FinerGit~\cite{higo2020tracking}, \texttt{git log}~\cite{Grund:2021}, and IntelliJ's \emph{show history for selection}~\cite{Grund:2021} approach. CodeShovel also exhibited high accuracy in independent evaluations by industry developers.  
Although a more recent tool, CodeTracker~\cite{jodavi2022accurate}, was shown to be more accurate than CodeShovel on a researchers-made oracle, CodeTracker was not independently validated with industry practitioners. Also, the run-time of CodeTracker is slower than CodeShovel, making it difficult to use for collecting the change history of 774,051 methods that we used. CodeShovel has been used successfully to perform method-level maintenance analysis (e.g.,~\cite{Chowdhury:2024:promise, Chowdhury:2022-deb, ahmad2024impact}).

CodeShovel offers an advanced method tracking system within Git repositories, enabling the capture of critical details about when, why, and how methods are altered over time. The output of CodeShovel is one JSON file for each method, containing all the change commits along with the different versions of source code and other important information. We developed a custom tool that interacts with CodeShovel, enabling us to retrieve comprehensive historical data and different code and change metrics for each method. The tool leveraged the popular JavaParser\footnote{https://github.com/javaparser/javaparser last accessed: Sep 20, 2024} library to parse the source code and generate abstract syntax trees (ASTs) that served as the primary source of code metric calculations. This tool was independently validated by two different researchers with 1000 randomly sampled Java methods. In all cases, the tool correctly measured different code metrics and change histories from the CodeShovel's output. 

\subsection{Detecting SATD in Methods}
\label{subsec:satd}
The objective of our study is to evaluate if SATD methods are more maintenance-prone than NOT-SATD methods. This required us to collect all the code comments contained in each of the studied methods. For a given method, we fed all of its comments to \emph{SATD detector}, a text-mining-based tool developed by Liu \textit{et al.}~\cite{Liu:2018}. We selected this tool because of its availability as an easy-to-use \texttt{jar} file and its adoption in other SATD studies~\cite{liu2021exploratory, zampetti2017recommending, bhatia2023empirical}. A source method was labeled as a method with SATD if at least one of its comments is identified as SATD by the \emph{SATD detector} tool.


\subsection{Visualization, Statistical Tests, and Design Choices}
\label{subsec:stat-normalization}
To compare various code quality and maintainability indicators between methods with and without SATD, we extensively use the Cumulative Distribution Function (CDF). Unlike the Probability Distribution Function (PDF), the CDF is monotonic, which makes it easier to interpret the data by avoiding irregular or zigzag patterns. Additionally, we aim to assess whether the visual differences between the SATD and NOT-SATD groups are statistically significant. 

Using the Anderson-Darling normality test~\cite{razali2011power}, we found that the distributions of our data did not follow a normal distribution. As a result, we employ the non-parametric Wilcoxon rank sum test to assess statistical significance. When we identify significant differences between distributions, we quantify the size of the effect using Cliff's $d$, a non-parametric measure of effect size. Cliff's $d$ ranges in the interval $[-1, 1]$ and is considered \textit{negligible} for $|d| < 0.147$, \textit{small} for $0.147 \leq |d| < 0.33$, \textit{medium} for $0.33 \leq |d| < 0.474$, and \textit{large} for $|d| \geq 0.474$.
Both the Wilcoxon rank-sum test and Cliff's delta are widely used in data-driven software engineering research~\cite{he2015empirical, chen2020savior,pecorelli2020testing,greenscaler:2019,Bangash}. Similar to other relevant studies~\cite{Inozemtseva:2014, Gil:2017, greenscaler:2019}, when calculating correlation coefficients, we used the non-parametric Kendall's $\tau$ instead of the Pearson correlation coefficient.

The methodological choices made by researchers in software maintenance studies impact the outcomes of their analysis~\cite{ahmad2024impact}. During our investigation into SATD-labeled methods and various maintenance indicators, we encountered two possible approaches for classification and analysis:

\begin{itemize} \item In the first approach, a method is labeled as a SATD method if it contains at least one SATD comment in any of its versions—ranging from its initial introduction to the most recent updates. Maintenance indicators (e.g., size, readability) can then be calculated based on different time points, such as the method’s introduction, its most recent state, or the mean or median values over its history. 

\item In the second approach, a method is labeled as SATD if it contains a SATD comment in its initial version. Subsequent maintenance indicators are calculated from later versions of the method. To qualify as a NOT-SATD method, the method must not contain any SATD comments throughout its entire lifetime. 
\end{itemize}

After thoroughly testing both approaches, we found that they did not substantially affect the study's results. We opted to present the findings based on the second approach, as it better aligns with the main objective of our research: it allows us to examine whether methods that start with SATD comments become more maintenance-prone over time.

\section{Approach, Analysis, and Results}
\label{results}
We now present the approach used to address each of the four research questions, along with the corresponding findings.

\subsection{RQ1: Do methods with SATD impact code quality?}
\label{subsec:RQ1}
Code metrics have been historically used as the reflective indicators of code quality and software maintenance~\cite{Gil:2017, Pascarella:2020, Chowdhury:2024:promise}. As such, we study the impact of SATD on code metrics measurements at the method level. We focus on 14 different code metrics from the following six broad categories. 

\textbf{\emph{Size.}} Size has been considered one of the most important code metrics to understand software maintenance burdens~\cite{Gil:2017, Chowdhury:2022:size, Emam:2001}. We calculated size as the number of source lines of code without comment and blank lines, in accordance with similar previous studies~\cite{Landman:2014,Chowdhury:2024:promise, Ralph:2018}.

\textbf{\emph{Readability.}} A core activity in software maintenance is code reading~\cite{Scalabrino:2016}. Code readability refers to how easily a developer can comprehend the structure and logic of the source code~\cite{Buse:2010, Posnett:2011, Borster:2016, Johnson:2019}. High readability is essential for efficient maintenance, as it reduces the cognitive load on developers and accelerates the process of updating and fixing code. In this study, we employed two distinct readability metrics: one proposed by Buse \textit{et al.}~\cite{Buse:2010} (\emph{Readability}) and the other by Posnett \textit{et al.}~\cite{Posnett:2011} (\emph{SimpleReadability}). Both metrics assign a readability score to each method, ranging from 0 (least readable) to 1 (most readable), providing a quantitative measure of how easily the code can be understood.

\textbf{\emph{Testability.}} We utilize the McCabe metric~\cite{McCabe:1976}, or cyclomatic complexity, to assess the testability of source methods. This metric quantifies the number of independent execution paths within a method, which serves as an indicator of its testing complexity. However, McCabe’s metric does not account for the number of control variables or comparison operations within a method's predicates. As such, we also incorporate two additional factors: the number of control variables (\emph{NVAR}) and the number of comparisons (\emph{NCOMP}), which contribute to a method’s overall complexity. Additionally, we include the McClure metric~\cite{McClure:1978}, which is calculated as the sum of \emph{NVAR} and \emph{NCOMP}. Calculating McCabe and McClure metrics typically requires a language-specific parser. Hindle \textit{et al.}~\cite{Hindle:2008} proposed proxy indentation as an alternative, which is also included in our study (\emph{IndentSTD}~\cite{Chowdhury:2022-deb}). Finally, recognizing that deep nesting can complicate testing, we also measured the Maximum Nested Block Depth.

\textbf{\emph{Dependency.}} If a method depends on too many other methods, any problems with those methods can be propagated to the caller method. Dependency, therefore, is an important quality and maintenance indicator. We calculated the total number of methods called by a given method (\emph{totalFanOut}) as its dependency measurement.  

\textbf{\emph{Maintainability.}} The Maintainability Index is a composite metric that combines various code metrics to provide a single score reflecting the maintainability of a software component. It is calculated as follows:
\begin{equation}
171 - 5.2 * ln(Halstead\, V) - 0.23 * (McCabe) - 16.2 * ln(Size)   
  \end{equation}
We include this index in our study because of its widespread adoption in industry-standard tools, such as Verifysoft Technology\footnote{\url{https://verifysoft.com/en_maintainability.html}: last accessed: Sep-28-2024} and Visual Studio\footnote{\url{https://docs.microsoft.com/en-us/visualstudio/code-quality/code-metrics-maintainability-index-range-and-meaning?view=vs-2022}: last accessed: Sep-28-2024}.

\textbf{\emph{Other.}} We also include Halstead Length, number of parameters, and number of local variables to answer this research question. We could not include the fan-in metric that we discuss in the threats to validity section. 

\subsubsection{Results}
\label{subsubsec:RQ2-Results}
\begin{figure*}[htbp]
\centering
\mbox{
\subfigure[]{\includegraphics[width=0.33\textwidth,keepaspectratio]{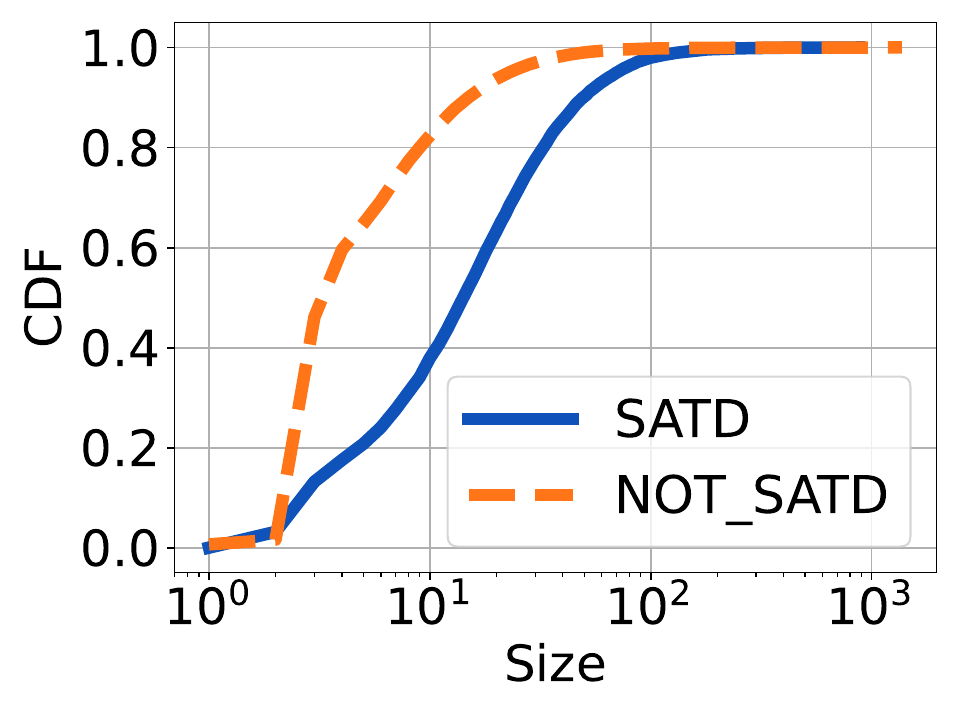}}
\subfigure[]{\includegraphics[width=0.33\textwidth,keepaspectratio]{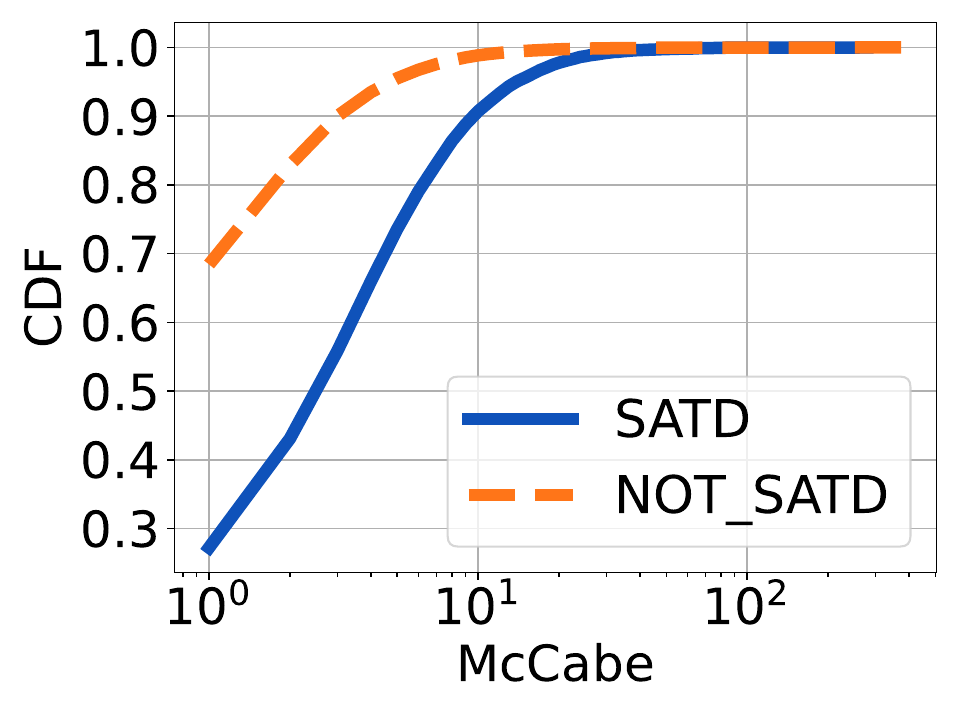}}
\subfigure[]{\includegraphics[width=0.33\textwidth,keepaspectratio]{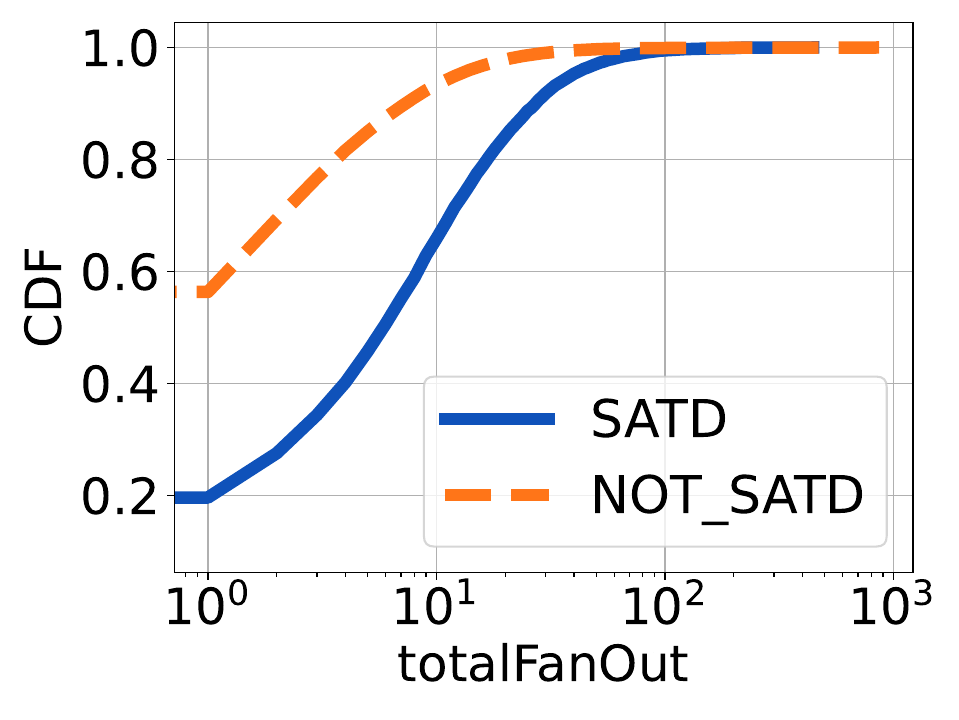}}
}

\mbox{
\subfigure[]{\includegraphics[width=0.33\textwidth,keepaspectratio]{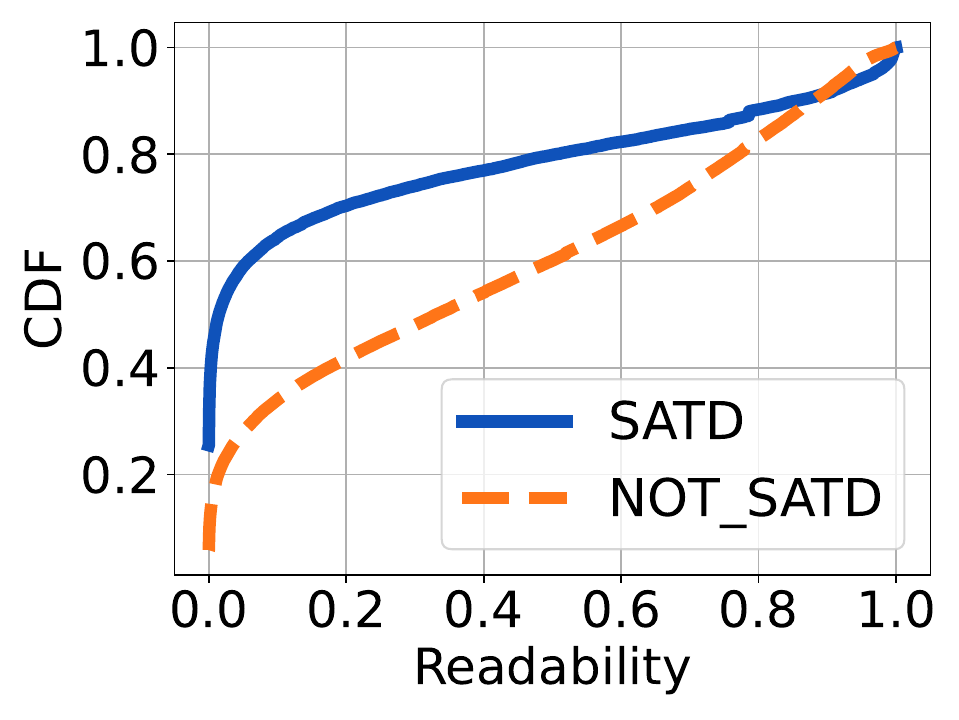}}
\subfigure[]{\includegraphics[width=0.33\textwidth,keepaspectratio]{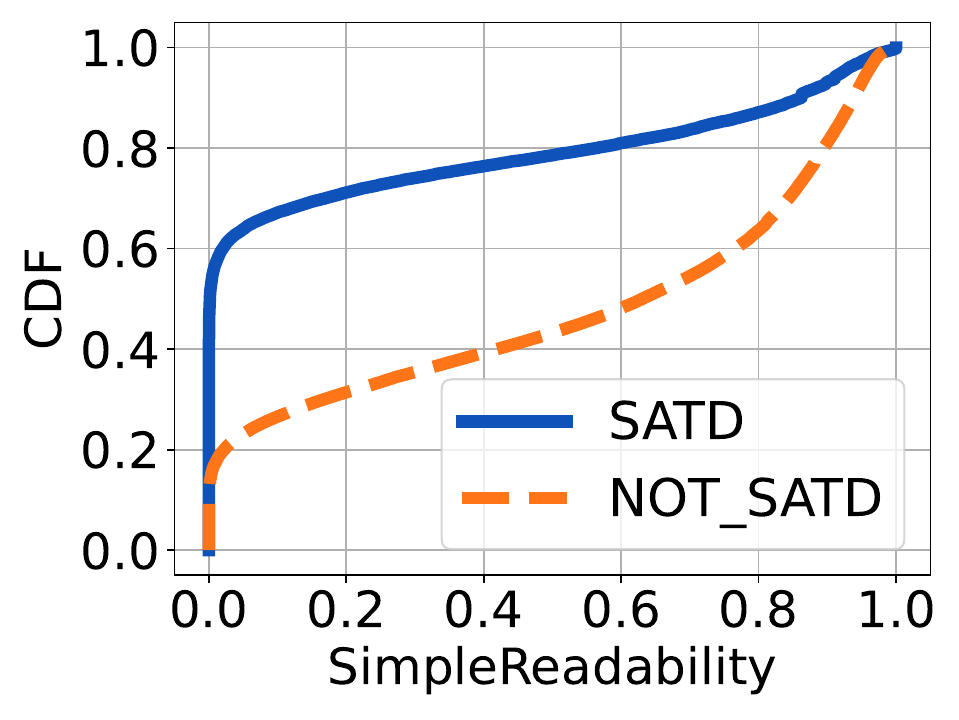}}
\subfigure[]{\includegraphics[width=0.33\textwidth,keepaspectratio]{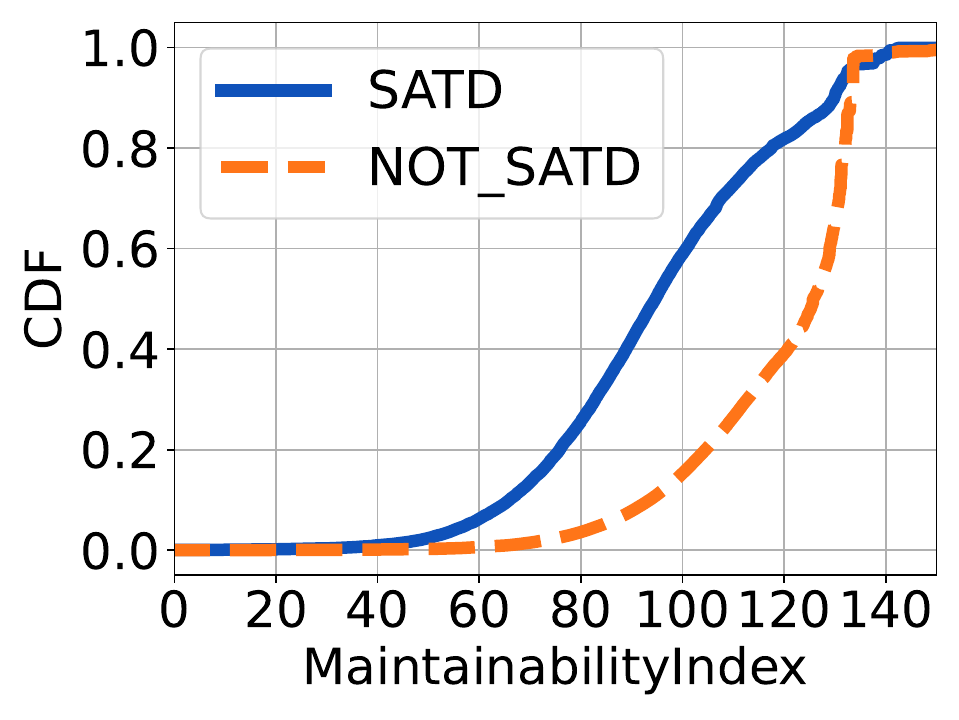}}
}
\caption{Differences in distribution for Size, McCabe, totalFanOut, Readability, SimpleReadability, and MaintainabilityIndex in the aggregated data. Clearly, SATD methods tend to have larger sizes, more complexity, more dependency, and lower readability, and worse Maintainability Index compared to NOT-SATD methods. 
} 
\label{fig:aggr-code-cdf}
\end{figure*}

Figure~\ref{fig:aggr-code-cdf} compares the distributions of six code metrics between SATD and NOT-SATD methods. To maintain clarity, we omit graphs for other metrics, though all of them exhibit patterns similar to the first row, with SATD methods showing larger values than NOT-SATD methods. 
These metrics were measured from the most recent versions of the methods to assess how SATD methods evolve compared to NOT-SATD methods. The results reveal a clear trend: SATD methods grow larger, less readable, and less maintainable over time. For instance, Figure~\ref{fig:aggr-code-cdf} (a) shows that 60\% of SATD methods have more than 10 lines of code, while only 20\% of NOT-SATD methods exceed this threshold. Similarly, the MaintainabilityIndex is above 100 for 80\% of NOT-SATD methods, compared to just 40\% for SATD methods.

We further investigate whether these differences are statistically significant and, if so, we calculate their effect sizes. The results for all 14 code metrics are shown in Table~\ref{tab:codemetrics-agg-stat}. The P-values from the Wilcoxon rank-sum test are all $\le 0.05$, indicating that the distributions of all code metrics are statistically significant. We also applied the Benjamini-Yekutieli correction approach~\cite{benjamini2001control} and found no noticeable difference in the results. 
Additionally, none of the effect sizes are small or negligible, except one; most are large, with a few exhibiting medium effect sizes. For three code metrics, marked with a `-' sign, the measurements for SATD methods are smaller than for NOT-SATD methods, reinforcing our earlier finding that SATD methods are less readable and maintainable.

To assess the generalizability of our results, we also conduct an analysis at the project level. The findings are summarized in Table~\ref{tab:codemetrics-indiv-stat}. Consider the \emph{Readability} metric as an example. We first determine the percentage of projects where the distribution difference between SATD and NOT-SATD methods is not statistically significant (i.e., $P > 0.05$). For projects where a statistical difference is found, we categorize the effect sizes as negligible, small, medium, or large. 
For most code metrics, differences between SATD and NOT-SATD methods are statistically significant across the majority of projects, with effect sizes predominantly falling between medium and large. The few projects that show no meaningful differences are those with a smaller number of SATD methods in our dataset, which limits the statistical power of the analysis.


\begin{table}[h!]
    \caption{Aggregated analysis for all the code metrics. The $P-values$ remain significant even after applying the Benjamini-Yekutieli correction approach~\cite{benjamini2001control}.}
    \centering
    \begin{tabular}{lccc}
    \hline
    \textbf{Metric} & \textbf{P} & \textbf{Sign} & \textbf{Effect Size} \\
    \hline
Size&0.00&+&large\\

Readability&0.00&-&medium\\

SimpleReadability&0.00&-&large\\

NVAR&0.00&+&medium\\

NCOMP&0.00&+&medium\\

Mcclure&0.00&+&medium\\

McCabe&0.00&+&large\\

IndentSTD&0.00&+&medium\\

MaximumBlockDepth&0.00&+&large\\

totalFanOut&0.00&+&large\\

Length&0.00&+&large\\

MaintainabilityIndex&0.00&-&large\\

Parameters&0.00&+&small\\

LocalVariables&0.00&+&large\\
    \hline
    \end{tabular}
    \label{tab:codemetrics-agg-stat}
\end{table}

\begin{table}[h!]
    \caption{Individual project analysis for all code metrics. The $P-values$ remain very similar after applying the Benjamini-Yekutieli correction.    
    Here, \textbf{N} refers to Negligible, \textbf{S} refers to Small, \textbf{M} refers to Medium, and \textbf{L} refers to Large effect size. For instance, for the size metric, the effect size is large for 81.63\% of projects.}
    \centering
    \begin{tabular}{lccccc}
    \hline
    \textbf{Metric} & \textbf{P $> 0.05$} & \textbf{N} & \textbf{S} & \textbf{M} &  \textbf{L}\\
    \hline
Size&0.00&0.00&4.08&14.29&81.63\\

Readability&6.52&2.17&8.70&43.48&45.65\\

SimpleReadability&2.08&0.00&8.33&20.83&70.83\\

NVAR&2.08&2.08&12.50&33.33&52.08\\

NCOMP&2.08&2.08&10.42&35.42&52.08\\

Mcclure&2.08&2.08&10.42&35.42&52.08\\

McCabe&4.26&0.00&10.64&25.53&63.83\\

IndentSTD&0.00&0.00&12.24&30.61&57.14\\

MaximumBlockDepth&4.26&0.00&10.64&27.66&61.70\\

totalFanOut&2.08&0.00&4.17&18.75&77.08\\

Length&0.00&0.00&2.04&8.16&89.80\\

MaintainabilityIndex&0.00&0.00&2.04&10.20&87.76\\

Parameters&28.95&2.63&63.16&34.21&0.00\\

LocalVariables&4.26&0.00&10.64&21.28&68.09\\

    \hline
    \end{tabular}
    \label{tab:codemetrics-indiv-stat}
\end{table}

\begin{tcolorbox}
    \textbf{\underline{RQ1 summary:}} Building on the longstanding assumption that code metrics are reliable indicators of software quality and maintainability, our findings confirm that SATD in source code significantly degrades both. Methods with SATD tend to grow larger, more complex, less readable, and less maintainable compared to methods that have never contained SATD.
\end{tcolorbox}

\subsection{RQ2: Do methods with SATD impact change-proneness?}
\label{subsec:RQ2}
Our previous analyses of code metrics have shown that SATD negatively impacts code quality and maintenance. However, there has been an ongoing debate in the literature regarding the true effectiveness of code metrics in understanding software maintenance challenges~\cite{Chowdhury:2022-deb}. For instance, while some studies have endorsed the utility of McCabe’s cyclomatic complexity~\cite{Curtis:1979, Landman:2014, Tiwari:2014, Zhou:2010, Alfadel:2017}, others have questioned its effectiveness~\cite{Weyuker:1988, Shepperd:1988, Gil:2017}. This skepticism extends to other widely-used code metrics as well~\cite{Gil:2017}.

\begin{figure*}[htbp]
\centering
\mbox{
\subfigure[]{\includegraphics[width=0.33\textwidth,keepaspectratio]{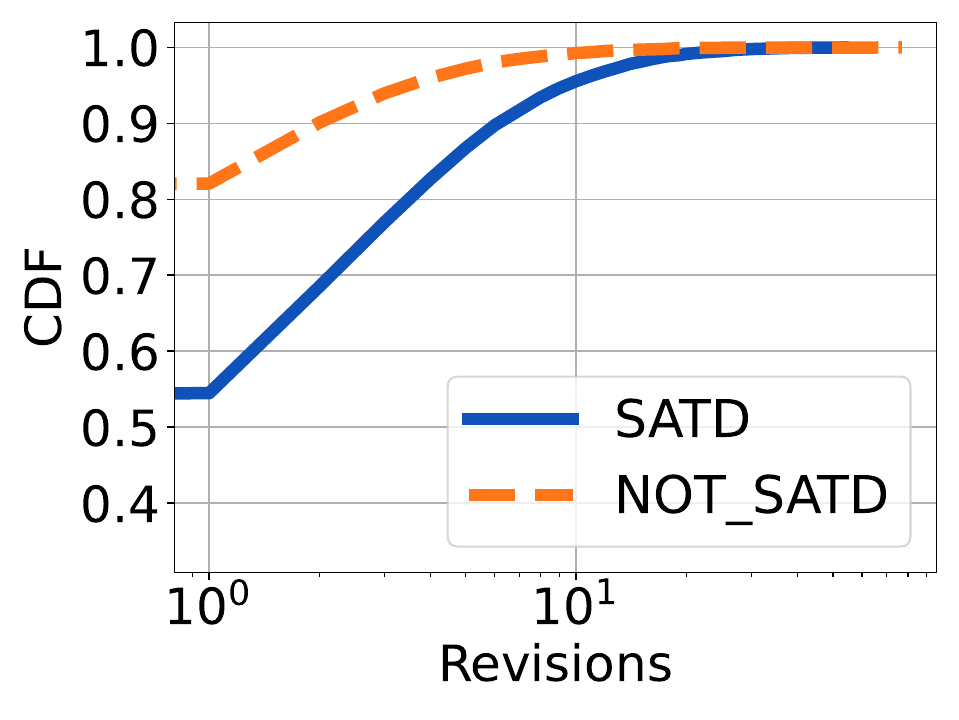}}
\subfigure[]{\includegraphics[width=0.33\textwidth,keepaspectratio]{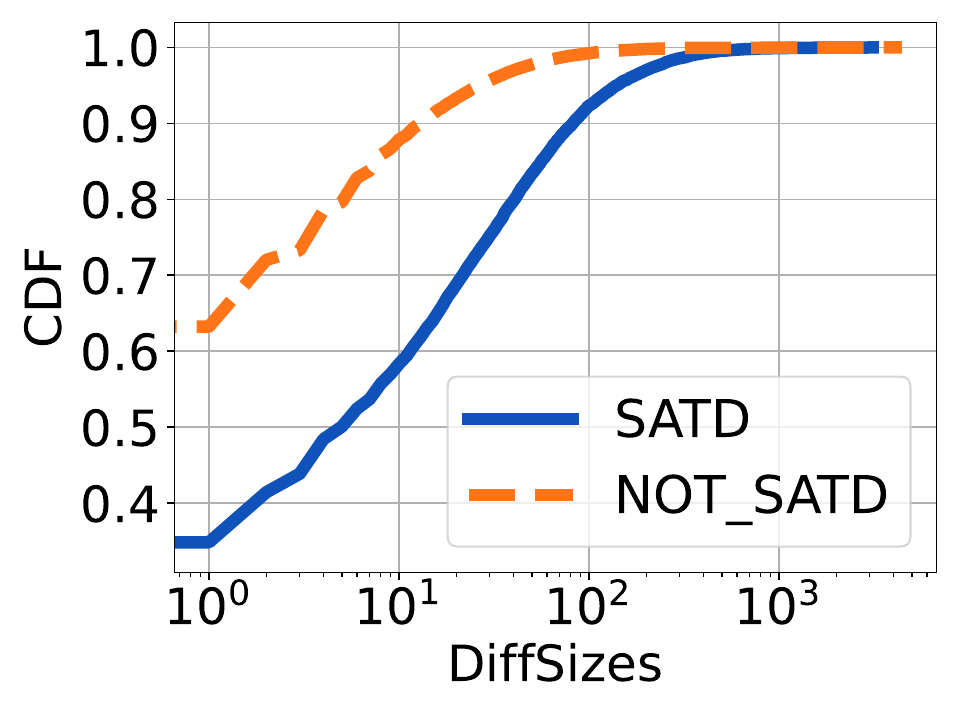}}
\subfigure[]{\includegraphics[width=0.33\textwidth,keepaspectratio]{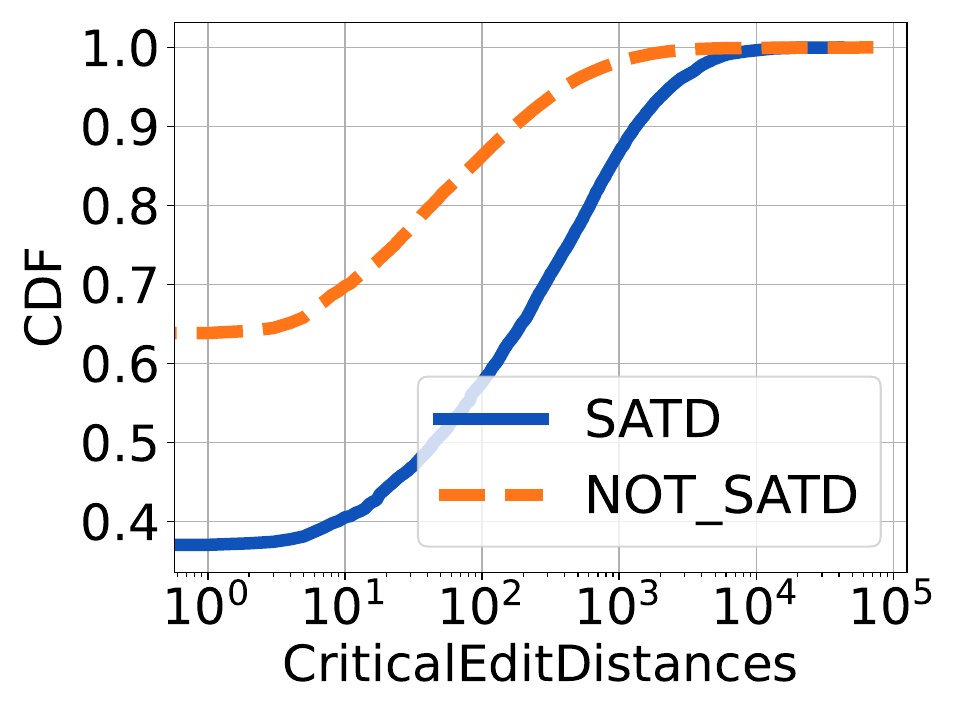}}
}

\caption{Differences in distribution for \#Revisions, DiffSizes, and CriticalEditDistances. Clearly, SATD methods tend to have more revisions and larger change sizes. The results are very similar for NewAdditions and EditDistances, therefore, not presented.} 
\label{fig:aggr-change-cdf}
\end{figure*}

Given these varying perspectives, we also focus on two well-established maintenance indicators---change- and bug-proneness---since the software engineering community generally agrees on their reliability for assessing maintenance burdens~\cite{arvanitou2017method, khomh2009exploratory, Khomh:2012, CATOLINO:2018, catolino2020improving, Romano:2011, Chowdhury:2024:promise, Pascarella:2020, Mo:2022}. In this RQ, we investigate how SATD influences the future change proneness of source methods. Specifically, we examine whether methods that start with SATD become more change-prone than methods that have never contained SATD. Encouraged by relevant previous studies, we selected the following five change-proneness indicators.

\textbf{\emph{\#Revisions.}}
The number of revisions a code component undergoes is often used as an indicator of its quality and maintainability, under the assumption that better-designed components require fewer modifications~\cite{Antinyan:2014, Monden:2002, Shin:2011, Antinyan:2015, Mo:2022, Pascarella:2020}. To calculate the revision count for a method, we relied on CodeShovel’s recorded change commits, excluding those where no actual code was altered---when a method is relocated or a file is renamed without changing its contents. 

\textbf{\emph{DiffSizes and NewAdditions.}}
The number of revisions alone does not fully reflect a component's change-proneness, as revisions can vary greatly in terms of the effort required—some may involve minimal changes, while others can be more extensive. To address this, diff size is often used as a more comprehensive measure~\cite{Mo:2022, Scholtes:2016, Shin:2011}. However, since adding new code is typically more challenging than removing it, the count of added lines (NewAdditions) is sometimes considered a better indicator of change-proneness~\cite{Shin:2011, Mo:2022, Pascarella:2020}. We calculated the DiffSizes by summing the diff sizes for each change commit. NewAdditions were computed in a similar fashion. This approach allowed us to capture the full scope of changes made to a method throughout its lifecycle, providing a more accurate measure of its evolution.

\textbf{\emph{EditDistances and CriticalEditDistances.}} DiffSizes and NewAdditions fail to account for the differences in the length of modified lines. To overcome this limitation, we employed the Levenshtein edit distance~\cite{levenshtein:1966, Stahl:2019, Scalabrino:2017, Scholtes:2016}, which quantifies the number of insertions, deletions, or substitutions required to convert one version of a method to another. Similar to DiffSizes and NewAdditions, we calculated the EditDistances by summing the edit distances for each change commit. CriticalEditDistances were computed in a similar manner, with the key difference that non-critical changes---such as whitespace modifications and comment updates---were excluded. This adjustment was necessary, as previous research has shown that such non-critical changes can skew the true impact of code modifications~\cite{ahmad2024impact}. 

\subsubsection{Results}
Figure~\ref{fig:aggr-change-cdf} illustrates that SATD methods experience more revisions and exhibit significantly larger change sizes compared to NOT-SATD methods. For example, more than 45\% of the SATD methods were revised at least once, which is true only for 20\% of NOT-SATD methods. Table~\ref{tab:change-agg-stat} further confirms that SATD methods are statistically distinct from NOT-SATD methods across all change-proneness metrics ($P \le 0.05$). Additionally, for all change-proneness indicators, SATD methods show medium effect sizes, with more revisions and larger change sizes, as indicated by the `+' sign.

\begin{table}[h!]
    \caption{Statistical Analysis for the aggregated data.} 
    \centering
    \begin{tabular}{lccc}
    
    \hline
    \textbf{Indicator} & \textbf{P} & \textbf{Sign} & \textbf{Effect Size} \\
    \hline
Revisions&0.00&+&medium\\

DiffSizes&0.00&+&medium\\

NewAdditions&0.00&+&medium\\

EditDistances&0.00&+&medium\\

CriticalEditDistances&0.00&+&medium\\
    \hline
    \end{tabular}
    \label{tab:change-agg-stat}
\end{table}

\begin{table}[h!]
    \caption{Individual project analysis. \textbf{N} refers to Negligible, \textbf{S} refers to Small, \textbf{M} refers to Medium, and \textbf{L} refers to Large effect size.  For instance, for the Revisions metric, the effect size is large for 61.70\% percent of projects.}
    \centering
    \begin{tabular}{lccccc}
    \hline
    \textbf{Indicator} & \textbf{P $> 0.05$} & \textbf{N} & \textbf{S} & \textbf{M} &  \textbf{L}\\
    \hline

Revisions&4.26&0.00&8.51&29.79&61.70\\

DiffSizes&2.08&0.00&6.25&20.83&72.92\\

NewAdditions&4.26&0.00&6.38&27.66&65.96\\

EditDistances&0.00&0.00&4.08&26.53&69.39\\

CriticalEditDistances&4.26&0.00&8.51&23.40&68.09\\

    \hline
    \end{tabular}
    \label{tab:change-indiv-stat}
\end{table}

We also investigated whether these change-proneness patterns vary across different projects. Table~\ref{tab:change-indiv-stat} presents the results for all projects. For every project, the EditDistances between SATD and NOT-SATD methods are statistically significant. In fact, the few projects where no statistical differences were observed for other metrics were primarily due to the limited number of SATD samples in those projects. Notably, the majority of the projects exhibit medium to large effect sizes.

Multiple studies~\cite{Chowdhury:2022-deb, ahmad2024impact} have shown that comparing methods of different ages when analyzing change- and bug-proneness can lead to inaccurate conclusions, due to the strong correlation between age and the likelihood of changes (or bugs). Older methods have had more time to evolve than newly introduced ones. Following the recommendation by Chowdhury \textit{et al.}~\cite{Chowdhury:2022-deb}, we applied a 2-year age normalization to assess whether it affected our findings. Specifically, based on the suggestion of Chowdhury \textit{et al.}~\cite{Chowdhury:2022-deb}, we excluded methods younger than 2 years and discarded any changes that occurred after the two-year mark. However, our analysis of change-proneness remained consistent, even after this age normalization. 

 \begin{tcolorbox}
     \textbf{\underline{RQ2 summary:}} Our findings, which take into account the number of revisions and various change sizes, clearly show that SATD methods are significantly more change-prone than NOT-SATD methods. 
 \end{tcolorbox}

\subsection{RQ3: Do methods with SATD impact code bug-proneness?}
\label{subsec:RQ3}

Bugs or defects in software are a major concern for stakeholders, often leading to significant challenges in maintenance and reliability. Given the importance of bug detection, the research community has extensively explored bug prediction and localization, with over 100,000 results on Google Scholar for the term \emph{bug prediction}. In this study, we investigate whether SATD methods are more likely to become bug-prone in the future compared to NOT-SATD methods. For each of the 49 projects, we calculated the ratio of bug-prone methods separately for SATD and NOT-SATD methods. This allows us to compare the bug-proneness between the two sets both at the aggregated level and for each project individually.

\begin{figure*}[htbp]
\centering
\mbox{
\subfigure[HighRecall]{\includegraphics[width=0.33\textwidth,keepaspectratio]{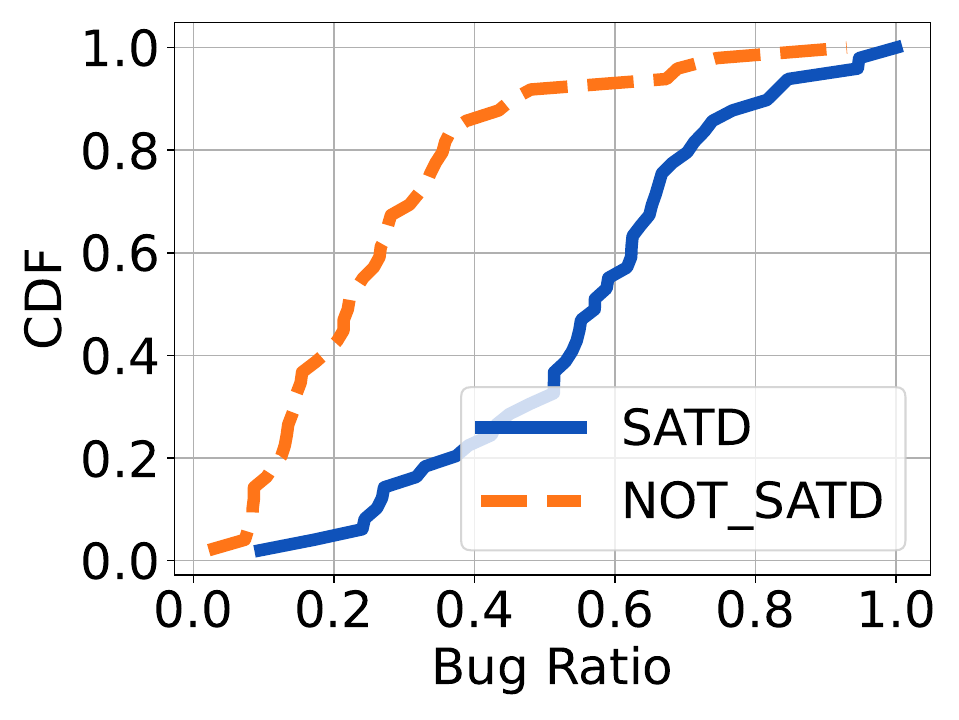}}
\subfigure[Balanced]{\includegraphics[width=0.33\textwidth,keepaspectratio]{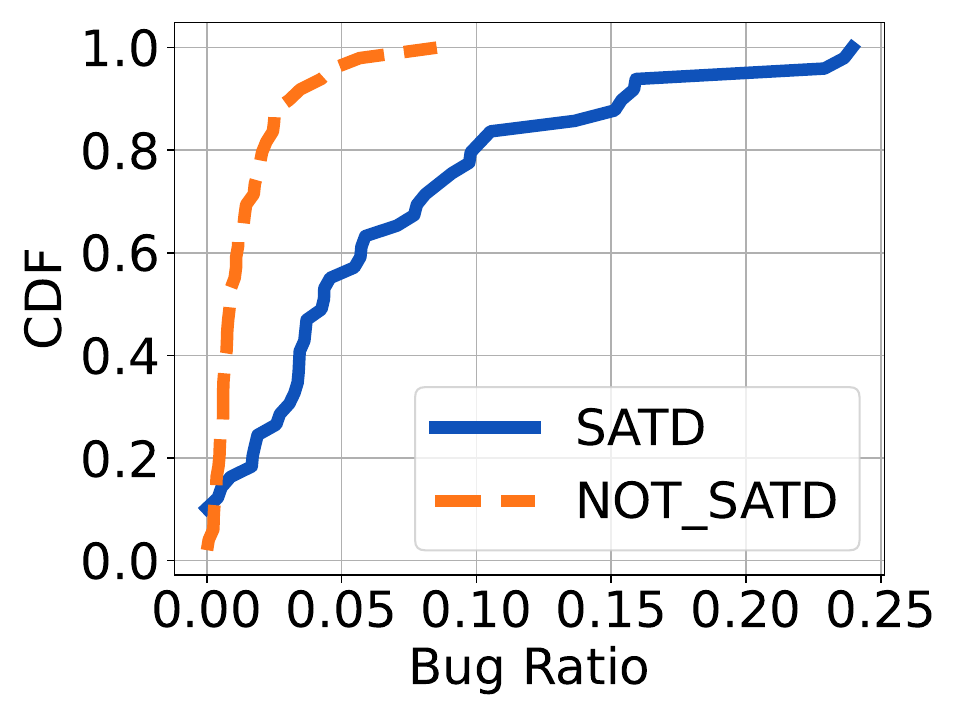}}
\subfigure[HighPrecision]{\includegraphics[width=0.33\textwidth,keepaspectratio]{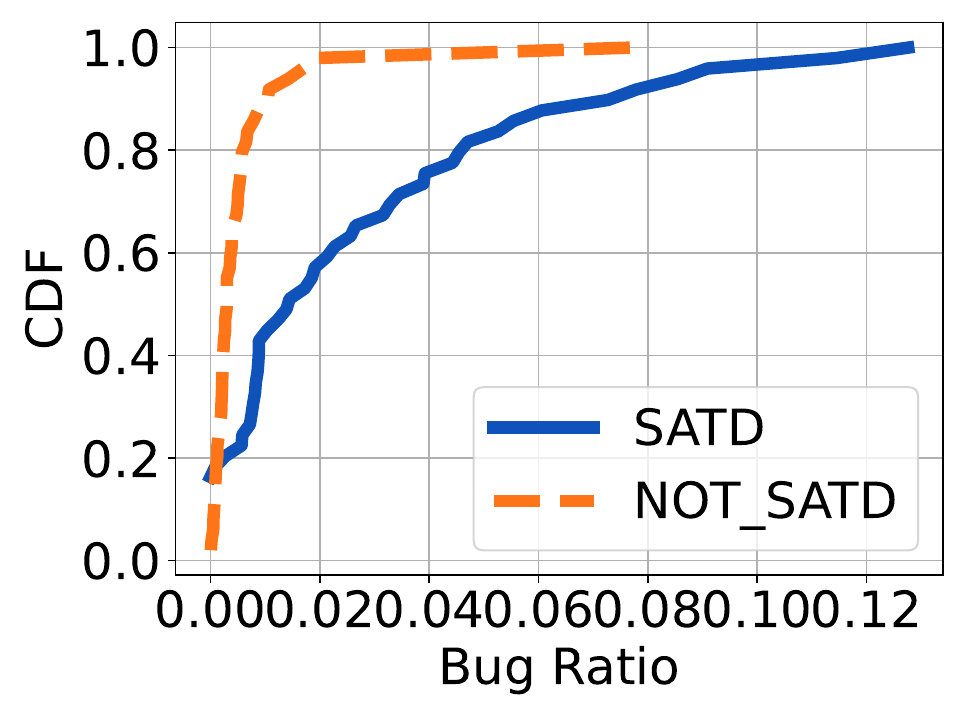}}
}

\caption{Differences in the distribution of bug ratios in three datasets: SATD methods tend to have more bugs than NOT-SATD methods in all datasets. } 
\label{fig:individual-bug-cdf}
\end{figure*}

A method is considered bug-prone if it is associated with a bug-fix commit at a later time. While it is common to use commit messages containing bug-fix keywords to identify bug-prone components~\cite{Mo:2022, Pascarella:2020, Chowdhury:2024:promise, Shippey:2016}, this approach is imprecise due to the issue of tangled changes. Unfortunately, tangled changes---unrelated modifications that are committed together---are common in the real world~\cite{kirinuki2014hey, Chowdhury:2024:promise, herzig2016impact}. 
To mitigate the impact of tangled changes on our analysis, we conduct our analysis on three different types of bug datasets. 
\begin{itemize}
    \item \textbf{HighRecall Dataset}. A method is a bug-prone method if it is modified in a commit containing one of the following nine keywords (or their variations): \emph{error, bug, fix, issue, mistake, incorrect, fault, defect, and flaw}. This set of keywords has been used in numerous studies~\cite{Ray:2016, Mocku:2000}. While this method is effective at capturing a broad range of buggy methods (resulting in high recall), it often leads to a significant number of false positives due to the challenges posed by tangled code changes.

    \item\textbf{HighPrecision Dataset}. To reduce false positives, Chowdhury \textit{et al.}~\cite{Chowdhury:2024:promise} refined the bug labeling approach by removing the keyword \emph{issue} and adding \emph{misfeature} based on the recommendation of Rosa \textit{et al.}~\cite{Rosa:2021}. They also included fix-related terms (\emph{fix, address, resolve}) to improve precision and excluded commits modifying more than one method to avoid tangled changes. This approach enhances precision but sacrifices recall by excluding many potential bug-fix commits.

    \item\textbf{Balanced Dataset}. We created this dataset using the same keyword set as the HighPrecision dataset, but with a key difference: instead of limiting bug-fix commits to modifications of a single method, we allowed up to five method modifications per commit. This adjustment aimed to strike a balance between precision and recall.     

\end{itemize}

\subsubsection{Results} Table~\ref{tab:bug-agg-stat} presents the bug ratios for SATD and NOT-SATD methods based on the aggregated data. As expected, the HighRecall dataset shows higher bug ratios in both sets compared to the other two datasets. However, in all three datasets, SATD methods consistently exhibit a significantly higher bug ratio than their NOT-SATD counterparts. 

Our findings from the aggregated data hold true across individual projects as well. Figure~\ref{fig:individual-bug-cdf} illustrates the distribution of bug ratios across all 49 projects for each of the three datasets. For instance, in the HighRecall dataset, SATD methods have a bug ratio exceeding 0.4 in 80\% of the projects. In contrast, only 18\% of the projects show a bug ratio greater than 0.4 for NOT-SATD methods. Table~\ref{tab:bug-indiv-stat} confirms that all visual differences in Figure~\ref{fig:individual-bug-cdf} are statistically significant ($P \le 0.05$), with effect sizes ranging from large to small. The `+' sign further indicates that higher bug ratios are typically associated with SATD methods.

 \begin{tcolorbox}
     \textbf{\underline{RQ3 summary:}} SATD methods are significantly more bug-prone than NOT-SATD methods, making them a significant liability in terms of software maintainability and reliability, increasing both cost and effort and harming reputation. 
 \end{tcolorbox}

\begin{table}[t]
    \caption{Bug ratio in aggregated data. }
    \centering
    \begin{tabular}{lccc}
    \hline
    \textbf{Bug Type} &\textbf{SATD methods} & \textbf{NOT-SATD methods} \\
    \hline
HighRecall&0.396&0.213\\
Balanced&0.035&0.011\\
HighPrecision&0.012&0.0046\\
    \hline
    \end{tabular}
    \label{tab:bug-agg-stat}
\end{table}

\begin{table}[t]
    \caption{Statistical Analysis for Individual Projects. The small effect size with the HighPrecison dataset is due to the smaller number of bug-prone methods in both SATD and NOT-SATD sets.}
    \centering
    \begin{tabular}{lccc}
    \hline
    \textbf{Bug Type} & \textbf{P} & \textbf{Sign} & \textbf{Effect Size} \\
    \hline
HighRecall&0.00&+&large\\
Balanced&0.00&+&medium\\
HighPrecision&0.00&+&small\\
    \hline
    \end{tabular}
    \label{tab:bug-indiv-stat}
\end{table}

\subsection{RQ4: How long does it take to resolve SATD?}
In RQ1, RQ2, and RQ3, we have demonstrated that SATD negatively impacts software quality and maintainability. This naturally leads to the question: \emph{how often are SATDs removed in real-world software projects, and what is the typical duration for their resolution?} While previous studies have explored these questions~\cite{maldonado2017empirical, liu2021exploratory}, they yielded significantly different results. Additionally, prior approaches considered an SATD removed if a method containing the comment was not found in a later phase. However, this is inaccurate, as methods can be renamed or moved to a different class or file. With the introduction of CodeShovel, which we used for change tracking, we are now able to capture such transformations, resulting in a more accurate dataset for tracking SATD removal.

\subsubsection{Results} 
\begin{figure*}[htbp]
\centering
\mbox{
\subfigure[Unresolved]{\includegraphics[width=0.25\textwidth,keepaspectratio]{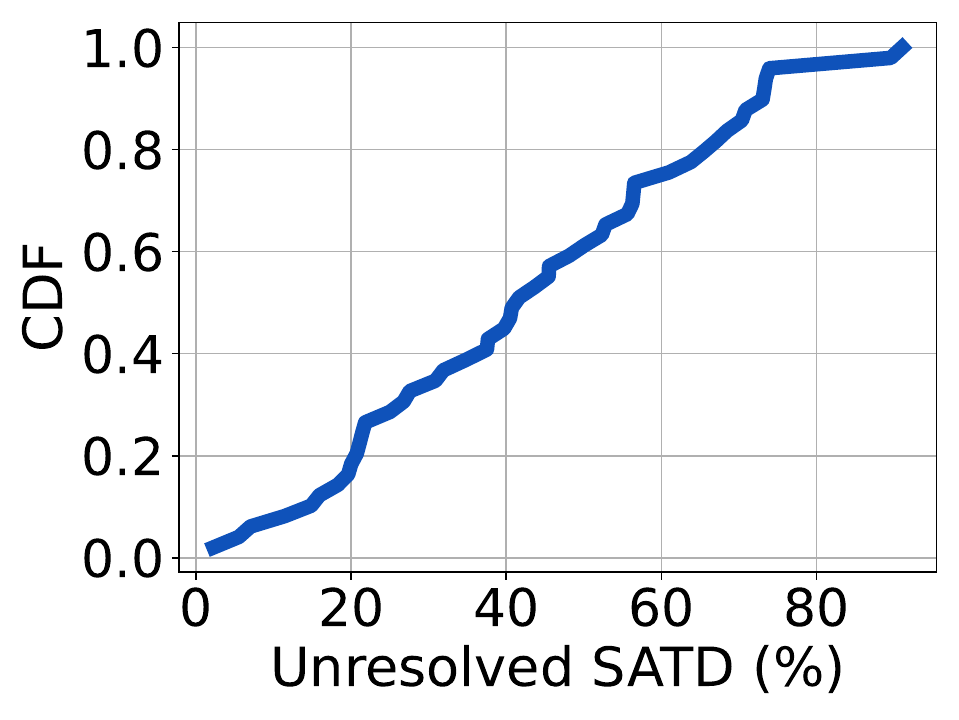}}
\subfigure[Aggregated]{\includegraphics[width=0.25\textwidth,keepaspectratio]{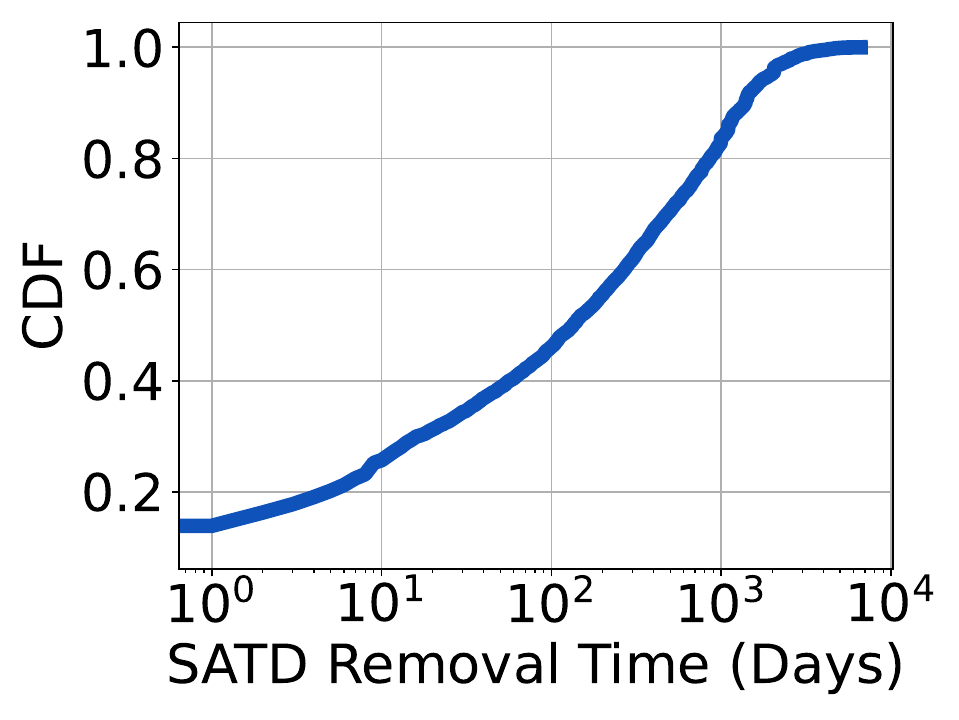}}
\subfigure[Mean]{\includegraphics[width=0.25\textwidth,keepaspectratio]{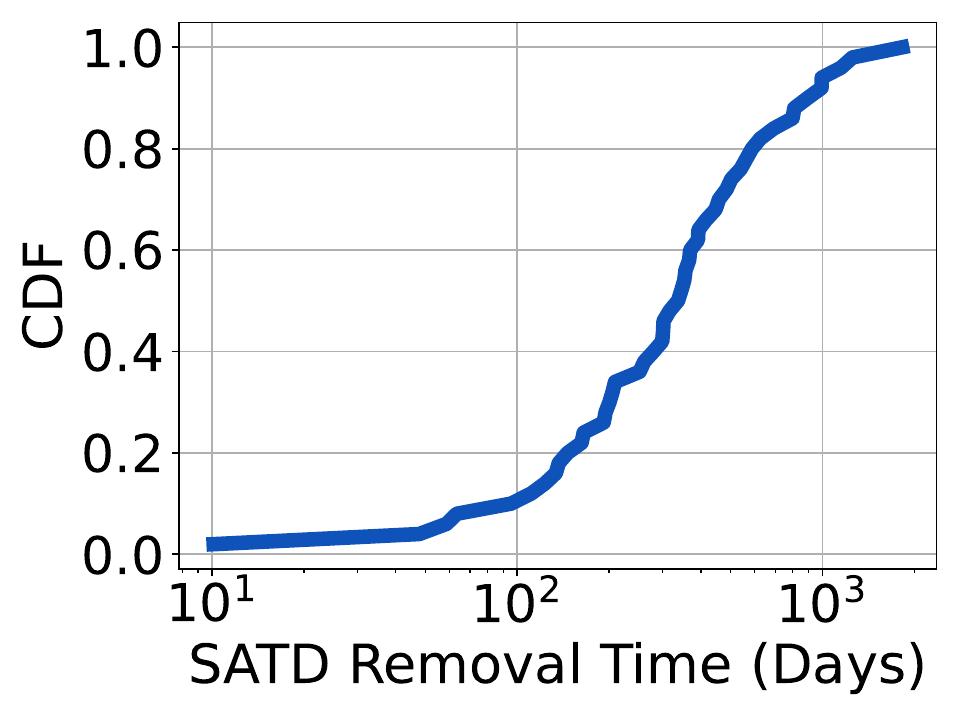}}
\subfigure[Median]{\includegraphics[width=0.25\textwidth,keepaspectratio]{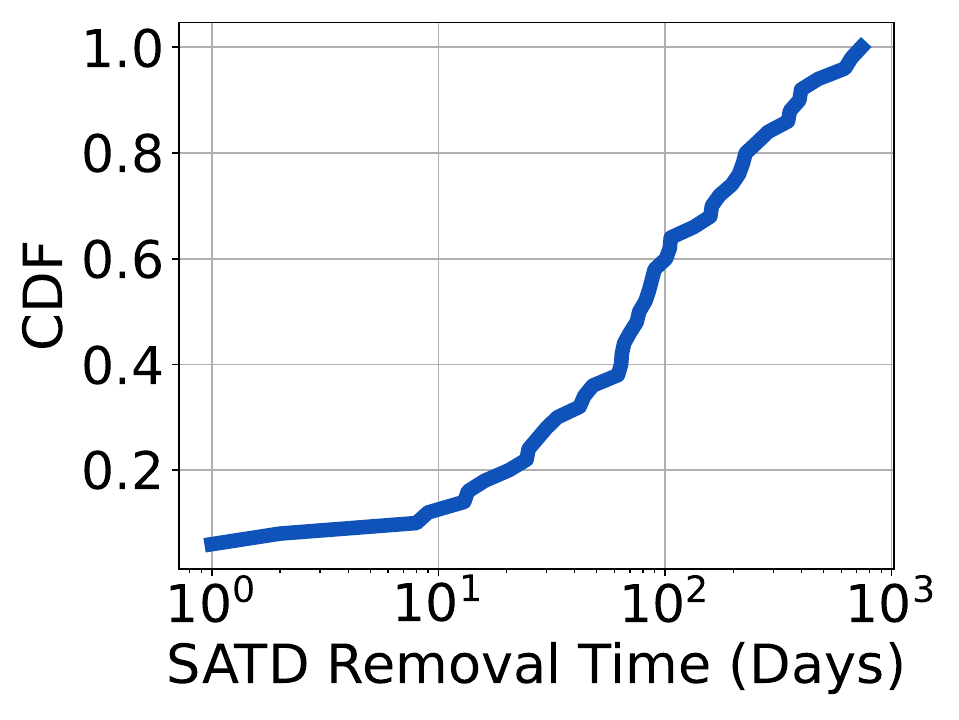}}

}

\caption{Figures (a) and (b) show the distribution of the percent of unresolved SATD methods and the distribution of SATD removal time in aggregated data, respectively. Figures (c) and (d) show the distribution of the means and medians of SATD removal time for all 49 projects.} 
\label{fig:resolution}
\end{figure*}

To ensure that newer SATD methods were not unfairly classified as unresolved, we excluded SATD methods younger than 2 years from our analysis, allowing them time to evolve and potentially address their SATDs. Nevertheless, our analysis of the aggregated data showed that over 61\% of SATDs were never removed (a similar observation was made by Zampetti \textit{et al.}~\cite{zampetti2018self}). The percentage of unresolved SATDs, however, varies considerably across different projects. Figure~\ref{fig:resolution} (a) shows the distribution of unresolved SATDs for all 49 projects. In 80\% of the projects, unresolved SATDs account for more than 20\%. Additionally, 20\% of the projects have unresolved SATDs exceeding 60\%, with a few projects showing more than 80\% unresolved SATDs.

For the resolved SATDs, we also calculated their removal times across the studied software projects. Figure~\ref{fig:resolution} (b) illustrates the distribution of removal times for all resolved SATDs in our dataset (aggregated data). We observe that 60\% of the SATDs took at least 100 days to be resolved, while 20\% required more than 1000 days to be resolved. 

To present the results for each project individually, we need to display 49 separate CDFs---one for each of the 49 projects---similar to the graph shown in Figure~\ref{fig:resolution} (b). To simplify, we computed the mean and median SATD removal times for each project. The distributions of these means and medians are shown in Figures~\ref{fig:resolution}(c) and~\ref{fig:resolution}(d), respectively. The results reveal significant variation in SATD removal times across projects. While some projects resolve SATDs relatively quickly, others take considerably longer, with mean and median removal times differing markedly between projects. For instance, approximately 10\% of projects have a median removal time of less than 10 days, while 40\% of projects experience a median removal time exceeding 100 days.

 \begin{tcolorbox}
     \textbf{\underline{RQ4 summary:}} A staggering $\sim$61\% of the SATDs are never removed from real-world software projects. When the SATDs are resolved, it may take more than 1000 days. While SATDs negatively impact all projects, the extent of their effect varies significantly between projects.
 \end{tcolorbox}

\section{Discussion}
\label{discussion}

Previous research has found no strong evidence supporting the harmful effects of Self-Admitted Technical Debts, yet considerable effort has been devoted to detecting and minimizing SATDs. In light of this, we conducted a comprehensive study to assess whether SATDs genuinely impact software quality and maintainability. Unlike prior studies~\cite{Potdar:2014, bavota2016large, wehaibi2016examining}, which focused on class/file level analysis, our research focused on method-level granularity---a level often preferred by both practitioners and researchers~\cite{Grund:2021, Pascarella:2020}.

Our findings demonstrate that SATDs are indeed detrimental to software quality and maintenance when assessed using commonly accepted maintenance and quality indicators. Specifically, we observed that methods containing SATDs tend to be larger, more complex, less readable, and harder to maintain \textbf{(RQ1)}. When examined through popular metrics, SATD methods were found to be significantly more change-prone \textbf{(RQ2)} and bug-prone \textbf{(RQ3)}, thereby escalating future maintenance burdens. Furthermore, our study reveals that the negative impact of SATDs can persist for extended periods, with many SATDs remaining unresolved \textbf{(RQ4)}.

\textbf{\emph{Implications for practitioners.}} The cost of software maintenance has been a long concern for industry practitioners~\cite{Kafura:1987, Borstler:2016}. Our study provides strong evidence that SATDs have a significant negative impact on software quality and maintenance. Therefore, practitioners should prioritize the detection and resolution of SATDs. By addressing SATDs early, teams can mitigate the long-term maintenance burden and reduce the risk of bugs and costly rework. 
SATDs, especially those that remain unresolved over time, may lead to sustained negative impacts on software maintainability. Practitioners should account for these long-term costs when making technical decisions. Rather than postponing the resolution of SATDs, it may be more cost-effective to address them promptly to avoid accumulating technical debt that could hinder future development or result in harder-to-fix bugs.  
Tools and metrics that evaluate software quality (e.g., code complexity, change-proneness, and bug-proneness) should consider including SATD as one of the indicators. This would allow practitioners to prioritize code segments containing SATDs, thus aligning technical debt management with the software's overall quality goals.

\textbf{\emph{Implications for researchers.}}
In this paper, we treated all types of SATDs as a single, undifferentiated category. However, previous research has shown that SATDs can be classified into several distinct types, including design, defect, documentation, requirement, and test debt~\cite{maldonado2015detecting}. Future research can replicate our study separately for each of these types and can answer whether these different SATD types impact software quality and maintenance efforts differently. Understanding these differences will enable developers to prioritize which types of SATDs require more immediate attention and which can be addressed with more flexibility. Additionally, such insights would allow researchers to develop targeted strategies for managing each type of SATD more effectively, optimizing the approach to technical debt resolution.

Given the strong negative impact of SATDs, our study should encourage the research community to build automated tools that can detect SATDs more accurately and provide suggestions for remediation. Research into such tools could help streamline the SATD management process, making it easier for practitioners to identify and address problematic code before it leads to more serious maintenance issues.

Moreover, based on our findings, researchers should consider incorporating SATD as a feature in machine learning models for bug and change prediction, if they have not already done so. We found that SATD has not been used (or underutilized) in such existing models~\cite{Pascarella:2020, Mo:2022,Chowdhury:2024:promise, Khomh:2012, CATOLINO:2018}.

\subsection{Threats to Validity}
\textit{External validity} is limited due to the selection of software projects in this study. The results from open-source software may not reflect the characteristics of closed-source software, and the findings may not be directly applicable to projects implemented in programming languages other than Java. However, the 49 selected projects are popular and have been frequently used in other code metrics studies. 

Although we intended to include the fan-in metric in RQ1, we were unable to do so due to the complexity of generating thousands of call graphs for each project. This challenge stemmed from the need to collect code metrics at different commits for the same software project. However, this limitation should not affect our conclusions, as SATD methods consistently showed significant differences from NOT-SATD methods across all other studied metrics. An exception in a single metric is unlikely to alter our overall findings.

\textit{Internal validity} may be influenced by our choice of statistical tests: the Wilcoxon Rank-Sum and Cliff's Delta tests. However, both tests are widely used and well-established in software engineering research. We identified SATD through source code comments, but this approach has limitations. In some cases, developers may not include comments when introducing SATD (e.g., they may add a commit message acknowledging SATD but omit a corresponding comment), or they might write a comment about SATD, remove the problematic code, but fail to update or remove the comment, leading to inconsistencies between the code and the comments. 

\textit{Construct validity} is hampered by our selection of the CodeShovel tool for constructing method change history. Also, our accuracy in detecting SATD and their removal can be impacted by developers' forgetfulness in writing or removing SATD comments. We also used the SATD detector tool developed by Liu \textit{et al.}~\cite{Liu:2018}. A recent study~\cite{sheikhaei2024empirical} has shown the potential of the Large Language Models (LLMs) in detecting SATD with higher accuracy.


\textit{Conclusion validity} of our study is influenced by the various threats outlined above.

\input
\section{Conclusion}
\label{conclusion}

In conclusion, when method level granularity is considered, our study provides strong evidence that Self-Admitted Technical Debts have a significant negative impact on software quality and maintainability. We found that methods containing SATDs are larger, more complex, harder to maintain, and more prone to changes and bugs, ultimately increasing the burden of future maintenance. This highlights the importance of early detection and resolution of SATDs, as unresolved debts can deteriorate software quality over time.

For practitioners, our findings emphasize the need to prioritize SATD management, not only to improve code quality but also to reduce long-term maintenance costs. We also call for further research exploring different types of SATDs and their varying impacts, as well as the development of automated tools for accurate detection and remediation. Additionally, future studies should consider integrating SATD into machine learning models for predictive maintenance.

Overall, our study underscores the critical role SATDs play in software maintainability, providing a solid foundation for future work focused on better understanding and managing technical debt at the method level.

%
\IEEEpeerreviewmaketitle

\section*{Acknowledgment}

The authors would like to thank the Undergraduate Research Awards (URA, University of Manitoba) and the University Research Grants Program (URGP, University of Manitoba) for supporting this research. Also, we have used ChatGPT to improve the presentation (e.g., checking grammar and spelling) of this manuscript.



%
\bibliographystyle{IEEEtran}
\bibliography{paper}

\end{document}